\documentclass[journal]{IEEEtran}
\usepackage{graphicx,hyperref}
\usepackage{amsfonts,amsmath}
\usepackage{cite}
\usepackage{subfigure}
\usepackage{xcolor}
\usepackage{algorithm,algpseudocode}

\begin{document}

 \title{Deep Learning-based Position-domain Channel Extrapolation for Cell-Free Massive MIMO}

\author{Jiajia~Guo,~\IEEEmembership{Member,~IEEE,}
         Chao-Kai~Wen,~\IEEEmembership{Fellow,~IEEE,}
         Xiao~Li,~\IEEEmembership{Member,~IEEE,}
        and~Shi~Jin,~\IEEEmembership{Fellow,~IEEE}\\
        Email: \{jiajiaguo, li\_xiao, jinshi\}@seu.edu.cn, chaokai.wen@mail.nsysu.edu.tw
 }

\maketitle

\begin{abstract}
 
To reduce channel acquisition overhead, spatial, time, and frequency-domain channel extrapolation techniques have been widely studied. In this paper, we propose a novel deep learning-based \underline{P}osition-domain \underline{C}hannel \underline{E}xtrapolation framework (named PCEnet) for cell-free massive multiple-input multiple-output (MIMO) systems. The user's position, which contains significant channel characteristic information, can greatly enhance the efficiency of channel acquisition. In cell-free massive MIMO, while the propagation environments between different base stations and a specific user vary and their respective channels are uncorrelated, the user's position remains constant and unique across all channels. Building on this, the proposed PCEnet framework leverages the position as a bridge between channels to establish a mapping between the characteristics of different channels, thereby using one acquired channel to assist in the estimation and feedback of others. Specifically, this approach first utilizes neural networks (NNs) to infer the user's position from the obtained channel. {The estimated position, shared among BSs through a central processing unit (CPU)}, is then fed into an NN to design pilot symbols and concatenated with the feedback information to the channel reconstruction NN to reconstruct other channels, thereby significantly enhancing channel acquisition performance. Additionally, we propose a simplified strategy where only the estimated position is used in the reconstruction process without modifying the pilot design, thereby reducing latency. Furthermore, we introduce a position label-free approach that infers the relative user position instead of the absolute position, eliminating the need for ground truth position labels during the localization NN training. Simulation results demonstrate that the proposed PCEnet framework reduces pilot and feedback overheads by up to 50\%.
\end{abstract}


\IEEEpeerreviewmaketitle

\section{Introduction}
\begin{figure}[t]
    \centering
    \includegraphics[width=0.8\linewidth]{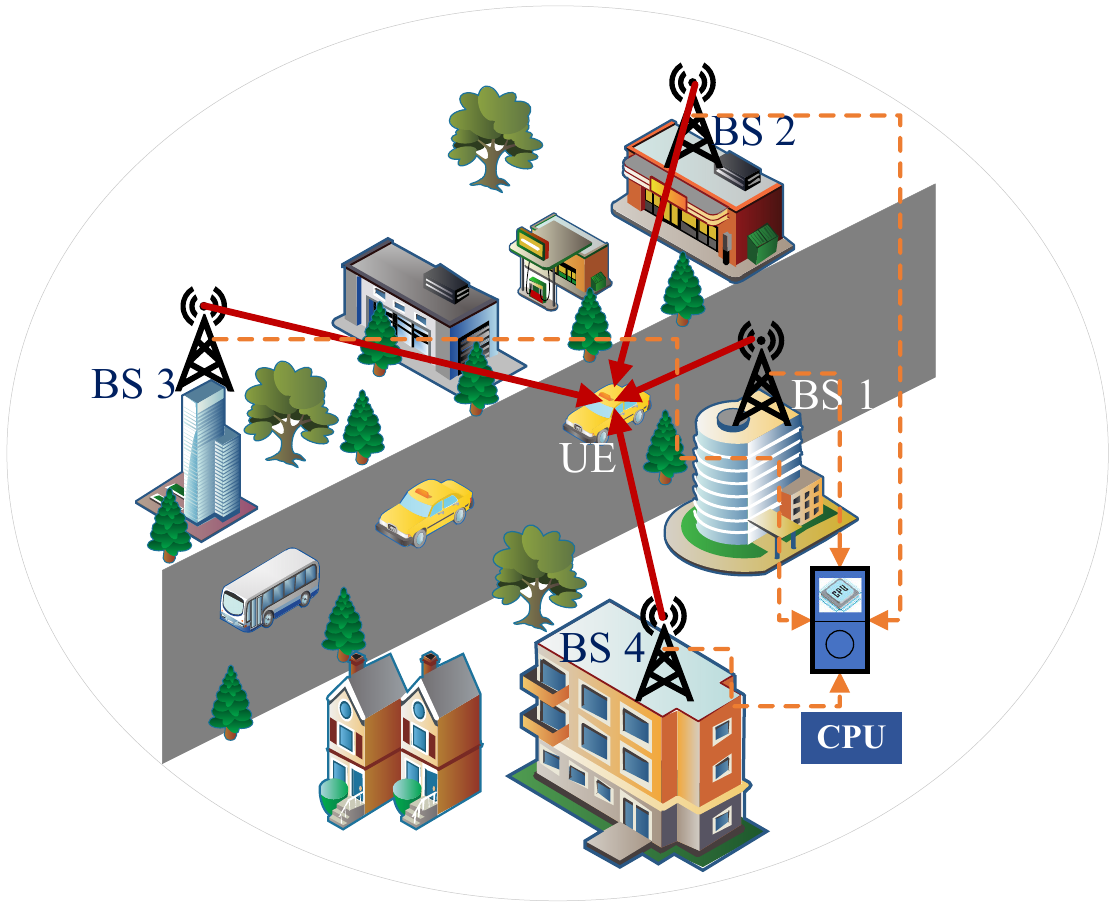}
     \caption{{An illustration of cell-free massive MIMO, a variant of massive MIMO, where multiple BSs, connected to a central processing unit (CPU), coherently serve multiple UEs over the same time-frequency resources.}} 
    \label{CFUSER}
    \vspace{-0.5cm}
\end{figure}

\IEEEPARstart{M}{assive} multiple-input multiple-output (MIMO), where the base station (BS) is equipped with numerous antennas to serve multiple user equipments (UEs) simultaneously \cite{5595728}, has significantly improved network performance and is a key technology in 5G \cite{10375688}. Looking ahead, massive MIMO is expected to remain central to 6G, with advancements such as extremely large-scale and cell-free massive MIMO being developed to enhance spectrum efficiency \cite{dang2020should,9390169,9170651,10379539,7827017}. In cell-free massive MIMO \cite{7827017}, large-scale antenna arrays, as illustrated in Figure \ref{CFUSER}, are formed through cooperation between multiple BSs to serve multiple UEs using the same time-frequency resources, resulting in significant performance gains.

The performance of both traditional and cell-free massive MIMO systems depends heavily on accurate channel state information (CSI) for efficient beamforming and interference management. As the number of antennas and the complexity of transmission topologies increase, the need for precise CSI becomes even more critical. In cell-free systems, inaccurate CSI can lead to poor beamforming and significant interference between BSs \cite{chen2022survey}. However, acquiring CSI incurs substantial overhead due to the need for pilot symbols and feedback \cite{10375688,10012511}. Reducing this overhead while maintaining accuracy remains a key challenge for 6G deployment. 

\subsection{Related Work}
The wireless channel is determined by the propagation environment and can be considered an electromagnetic fingerprint of that environment. Notably, propagation environments across different domains exhibit inherent correlations. As a result, {channel extrapolation, which predicts unobserved CSI using available CSI by leveraging cross-domain correlations (e.g., in time, frequency, or space) \cite{10024904}, has been widely studied as a promising approach to reduce or even eliminate CSI acquisition overhead in both time-division duplexing (TDD) and frequency-division duplexing (FDD) massive MIMO systems \cite{10403774,8968763}.} Existing research on channel extrapolation can be categorized into three main approaches \cite{9427294,10024904}:

{\textbf {Time-domain channel extrapolation:}}
Over short time periods, the propagation environment changes minimally, resulting in highly correlated channels. Studies \cite{1512123,8979256} use conventional or learning-based algorithms to predict future channels from past data, addressing channel aging and potentially eliminating the need for channel estimation and feedback. Other works \cite{7390019,8482358} use past channels to improve future estimation and feedback efficiency. 

{\textbf {Frequency-domain channel extrapolation: }}
Signals transmitted over different subcarriers experience the same propagation environment, leading to strong frequency-domain correlations. In FDD systems, uplink and downlink channels exhibit partial reciprocity (e.g., angular reciprocity) \cite{9322570,10345764}, which can be exploited to enhance downlink training and feedback efficiency \cite{8968763}. Some studies propose directly predicting downlink CSI from uplink CSI, thereby eliminating the need for additional pilot signals and feedback \cite{2934895,9048929,3649343}.  

{\textbf {Spatial-domain channel extrapolation:}}
Antennas in close proximity experience similar environments, resulting in correlated spatial-domain channels. { In large antenna arrays, the channel for the entire array can be extrapolated from a subset of antennas \cite{9048929,10423003,10566602}.} Additionally, angular correlations among UEs sharing the same scatterers facilitate CSI acquisition, enabling techniques such as joint CSI reconstruction at the BS \cite{8954618}. 

{ These channel extrapolation approaches are not mutually exclusive but can be combined to further improve CSI accuracy while reducing pilot and feedback overhead in massive MIMO systems \cite{9427294}. For example, spatial and temporal channel extrapolation have been jointly applied in \cite{9857844} using variational autoencoders.}
Channel extrapolation has also gained industry attention. In both TDD and FDD systems, the 3rd Generation Partnership Project (3GPP) Releases 16, 17, and 18 have incorporated channel correlations across time, frequency, and spatial domains by identifying sparse bases to approximate the full beamforming matrix, thereby enhancing CSI feedback efficiency \cite{9904665,10121037}.

\subsection{Main Focus of This Work}
\label{s1b}
In cell-free massive MIMO, large-scale antenna arrays are constructed through cooperation among multiple BSs. While traditional channel extrapolation techniques exploit domain-specific correlations, they rely on common propagation environments and overlook the distinctive structure of cell-free systems---namely, the distributed antenna array layout. This raises a fundamental question:

\begin{center} \emph{Can channel extrapolation be effectively applied across different propagation environments, leveraging the unique structure of cell-free systems, to enhance CSI acquisition performance?}
\end{center}
To address this question, our study explores a deep learning-based \underline{\bf P}osition-domain \underline{\bf C}hannel \underline{\bf E}xtrapolation framework (named PCEnet) for cell-free massive MIMO, aiming to further enhance channel acquisition, going beyond the conventional frequency, time, and spatial domains.

In existing works \cite{9014542}, due to the distinct propagation environments between multiple BSs and the UE caused by the different transmitter locations\footnote{The distance between transmitters is much larger than in previous studies, which considered distances of 1 to 5 meters \cite{10423003}.}, the corresponding channels\footnote{In the following sections, the link between the UE and a specific BS is referred to as the main channel, while the links to other BSs are referred to as side channels. The terms ``main channel'' and ``side channel'' are used here for descriptive convenience and do not reflect their relative importance.} are uncorrelated and acquired separately in cell-free massive MIMO \cite{9014542}.
While main and side channels in cell-free systems exhibit no direct correlations, the same UE position hints at an underlying connection between them, and some prior works have explored directly inferring side channels from the main channel in a ``black-box'' manner, thereby eliminating channel acquisition overheads \cite{9048929}. The distance between BSs in this simulation is a mere 0.5 meters, significantly shorter than the distances typically found in practical cell-free systems. 

In contrast, we propose a ``white-box'' approach that explicitly estimates the UE's position and utilizes it to bridge the main and side channels, thereby reducing (rather than eliminating) CSI acquisition overhead. Our main contributions are summarized as follows: 

\begin{itemize}
    \item {\textbf{Position-domain channel extrapolation:}}
    We introduce an end-to-end CSI acquisition framework that employs deep neural networks (NNs) for pilot design, pilot signal feedback, and channel reconstruction, eliminating the need for explicit CSI estimation at the UE. Once the BS obtains the main channel, it estimates the UE's position using NNs. This estimated position is then used to optimize pilot design for the side channel, and the UE directly feeds back the received pilot signals to the BS. The BS then reconstructs the side channel using the feedback and estimated position, reducing pilot and feedback overhead by approximately 50\%.

    \item {\textbf{One-sided real-time channel extrapolation:}}
    In latency-sensitive scenarios, redesigning pilot signals for the side channel may introduce intolerable delays. To address this, we propose an alternative approach that retains the original pilot design while incorporating estimated position information into the CSI reconstruction process at the BS. { Since additional computations occur only at the BS, this approach is termed ``one-sided.''} While it may not be as optimal as the full framework, it still reduces pilot and feedback overhead by approximately 33.3\%.

    \item {\textbf{Position label-free channel extrapolation:}} 
    Training position-based NNs requires labeled position data, which can be challenging to obtain in real-world deployments. To circumvent this, we propose an autoencoder-based approach that learns a relative position representation instead of absolute position labels. This method reduces CSI acquisition overhead by more than one-third, even without ground-truth position labels.

\end{itemize}

The remainder of this paper is organized as follows: Section \ref{s2} presents the system model, including the massive MIMO and channel acquisition models. Section \ref{BenchmarkFrameworkSec} introduces a benchmark learning-based end-to-end CSI acquisition framework, followed by the proposed PCEnet framework and its two variants in Section \ref{PCEnetSection}. Section \ref{s4} describes the simulation settings and presents the results. Finally, Section \ref{s5} concludes the paper.

\section{System Model}
\label{s2}

\subsection{Cell-free Massive MIMO Model}
This study considers a narrowband downlink cell-free massive MIMO system consisting of $M$ BSs, each equipped with a uniform linear array (ULA) of $N$ transmitting antennas\footnote{{
The proposed method is applicable to both FDD and TDD systems. The pilot design, channel estimation, and feedback considered in this study are essential for practical implementations, as FDD lacks channel reciprocity \cite{10345764}, while TDD suffers from imperfect channel reciprocity \cite{8746703}.
}}. {All BSs are connected to a CPU to facilitate information exchange}, and together, the $M$ BSs simultaneously serve a single UE with a single receiving antenna over the same time-frequency resources.
The received signal $y\in \mathbb{C}$ at the UE can be expressed as 
\begin{equation}   
    y = \left( \sum_{i=1}^M {\bf h}_{i} {\bf v}_i \right) x + n,
\end{equation}
where ${\bf h}_{i} \in \mathbb{C}^{1\times N}$ represents the channel between the UE and the $i$-th BS, ${\bf v}_i\in \mathbb{C}^{ N\times 1}$ denotes the precoding vector of the $i$-th BS, and $x\in \mathbb{C}$ and $n\in \mathbb{C}$ represent the transmitted symbol and complex additive white Gaussian noise (AWGN), respectively.

Let ${\bf h}_{\rm all} = [{\bf h}_{1},\dots, {\bf h}_{M} ] \in  \mathbb{C}^{1\times MN}$ and ${\bf v}_{\rm all} = [{\bf v}_0^T,\dots,{\bf v}_M^T ]^T\in  \mathbb{C}^{ MN\times 1}$ be the stacked channel and precoding vectors, respectively. As illustrated in Figure \ref{CFUSER}, the CPU designs the precoding vector ${\bf v}_{\rm all}$ based on the acquired channel information ${\bf h}_{\rm all}$.
The quality of channel ${\bf h}_{\rm all}$ is crucial for the precoding performance. In cell-free massive MIMO systems, the number of BSs $M$ and the number of antennas per BS $N$ are typically large. This high scalability necessitates longer pilot sequences for accurate channel estimation and leads to significant overhead in the channel feedback process.

\subsection{Channel Acquisition Model}
Due to the distinct signal propagation environments between the UE and each BS, there is a lack of correlation among the respective channels $[{\bf h}_{1},\dots, {\bf h}_{M} ]$. Consequently, current methods treat the estimation and feedback of each channel independently.
In this study, we define the channel between the first BS and the UE as the main channel (i.e., ${\bf h}_{1}$), denoted by ${\bf h}_{\rm m}$, and the channel between the second BS and the UE (i.e., ${\bf h}_{2}$) as the side channel, denoted by ${\bf h}_{\rm s}$.\footnote{While ${\bf h}_{2}$ is used as an example, the proposed methodology can be extended to other channels, i.e., ${\bf h}_{3},\dots, {\bf h}_{M}$.} Correspondingly, we refer to the transmitters in the main and side channels as the main and side BSs, respectively.

To estimate the main channel, pilot signals ${\bf X}_{\rm m} \in  \mathbb{C}^{N  \times L_{\rm m}}$ of length $L_{\rm m}$ ($\ll N$) are transmitted to the UE, and the received signal ${\bf y}_{\rm m} \in  \mathbb{C}^{1\times L_{\rm m}}$ at the UE can be written as
\begin{equation}
\label{pilotTrans}
    {\bf y}_{\rm m} = {\bf h}_{\rm m}{\bf X}_{\rm m} + {\bf z}_{\rm m},
\end{equation}
where each pilot transmission satisfies the power constraint $P$, and ${\bf z}_{\rm m}\in  \mathbb{C}^{1\times L_{\rm m}}$ represents the AWGN  vector. The UE estimates the main channel ${\bf h}_{\rm m}$ using the known pilot ${\bf X}_{\rm m}$ and the received signal ${\bf y}_{\rm m}$.
For example, in least squares (LS) channel estimation, the estimation problem can be formulated as
\begin{equation}
\label{LSchannel}
    \mathop{\arg\min}\limits_{\widehat{\bf h}_{\rm m}} \|{\bf y}_{\rm m} -\widehat{\bf h}_{\rm m} {\bf X}_{\rm m}\|_2^2,
\end{equation}
where $\|\cdot\|_2$ represents the Euclidean norm, and $\widehat{\bf h}_{\rm m}$ is the estimated main channel. Once the UE estimates $\widehat{\bf h}_{\rm m}$, it feeds this information back to the main BS.

To reduce channel estimation overhead, the pilot sequence length ${\bf X}_{\rm m}$ is often much smaller than the number of antennas $N$, i.e., $L_{\rm m} \ll N$. However, this increases estimation errors, requiring more advanced channel estimation techniques. Similarly, to minimize feedback overhead, the estimated channel is often compressed before feedback using methods such as compressive sensing \cite{8350399}, codebooks \cite{3gpp214}, or deep learning \cite{9931713}.
For instance, in a deep learning-enabled CSI feedback system based on an autoencoder \cite{10639525}, the encoder at the UE compresses the estimated channel ${\widehat{\bf h}_{\rm m}}$ using an NN function  ${\rm f}_{\rm en}(\cdot)$ with parameters ${\bf \Theta}_{\rm en}$.
The decoder at the main BS then reconstructs the compressed channel  ${\bf s}_{\rm m}$ using an NN function ${\rm f}_{\rm de}(\cdot)$ with parameters ${\bf \Theta}_{\rm de}$.
These operations are represented as
\begin{align}
    {\bf s}_{\rm m} &= \mathcal{Q}\big(  {{\rm f}_{\rm en}}({\widehat{\bf h}_{\rm m}};{\bf \Theta}_{\rm en}) \big), \\
      \widehat{\bf h}'_{\rm m} &= {{\rm f}_{\rm de}}({\bf s}_{\rm m};{\bf \Theta}_{\rm de}), \label{BSrec}
\end{align}
where $\mathcal{Q}(\cdot)$ represents the quantization operation,  ${\bf s}_{\rm m}$ is the compressed channel, and $\widehat{\bf h}'_{\rm m}$ is the reconstructed channel estimate.

The acquisition of the side channel ${\bf h}_{\rm s}$ follows a similar process as the main channel ${\bf h}_{\rm m}$, which is omitted here for brevity.

\begin{figure}[t]
    \centering
    \includegraphics[width=0.99\linewidth]{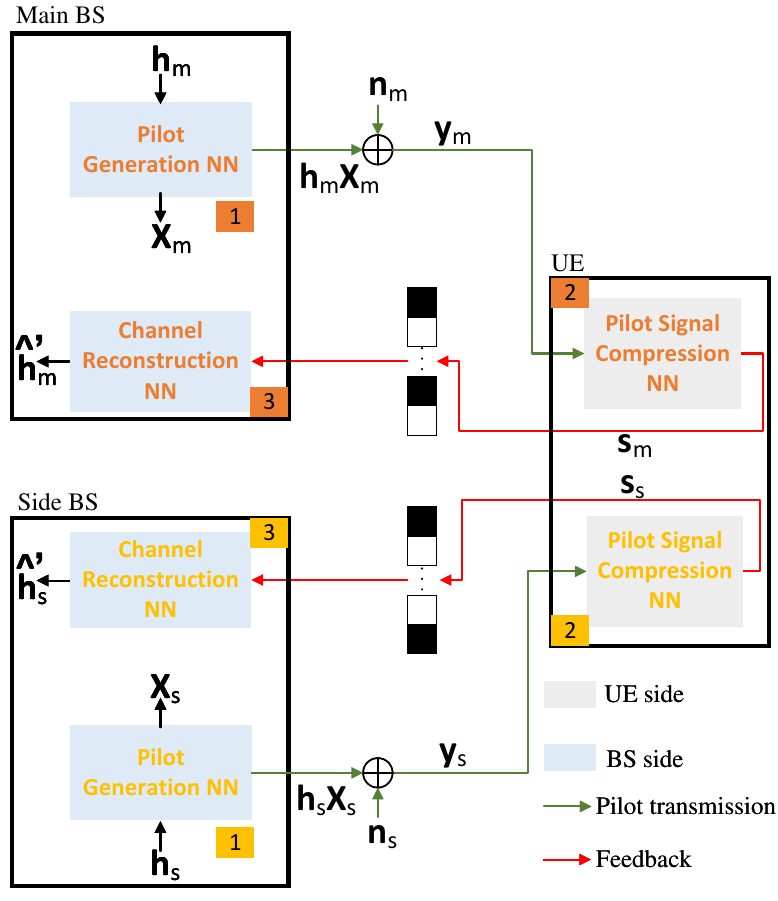}
    \caption{{Illustration of the end-to-end CSI acquisition framework, E2E-AI4CSI, including pilot design, pilot signal compression, and channel reconstruction.}}
    \label{E2Ebenchmark}
\end{figure}

\section{E2E-AI4CSI: Benchmark End-to-End CSI Acquisition Framework}
 
\label{BenchmarkFrameworkSec}

In this section, we present an end-to-end CSI acquisition framework, which serves as a benchmark NN model. This framework provides a foundation for comparison with the proposed PCEnet framework. 

\subsection{Motivation}
Equations (\ref{pilotTrans})--(\ref{BSrec}) outline the entire CSI acquisition process, including pilot design, transmission, channel estimation, and feedback. These components are interconnected and influence one another. However, optimizing each module individually does not guarantee global optimality across the entire system. For instance, regardless of how accurately the channel is estimated, poor feedback quality or limited feedback capacity can degrade the overall CSI quality at the BS. Therefore, the entire CSI acquisition process should be jointly designed for optimal performance.
 
End-to-end learning with deep learning offers a powerful approach to this challenge, as it allows for the seamless integration of multiple modules into a joint design \cite{9446676}. Joint pilot design and channel estimation have been explored in \cite{9037126}, while joint channel estimation and feedback are investigated in \cite{9785820}. Additionally, \cite{9570376} proposes a comprehensive end-to-end framework for uplink-aided downlink CSI acquisition. The results from these studies demonstrate that an end-to-end design significantly improves CSI acquisition accuracy and reduces overhead. 

\subsection{Main Framework}
Inspired by these advances, this study adopts a deep learning-based {\underline{end-to-end}} {\underline{CSI}} acquisition framework as the benchmark. The framework, named E2E-AI4CSI, is depicted in Figure \ref{E2Ebenchmark}. The key distinction of the E2E-AI4CSI framework from conventional methods lies in the direct feedback of the received pilot signal, eliminating the need for channel estimation at the UE.
In this process, we use the acquisition of the main channel (${ \bf h}_{\rm m}$) as an example. The main BS transmits a pilot symbol (${\bf X}_{\rm m}$) designed by the NNs to the UE. The UE compresses and quantizes the received pilot signal (${ \bf y}_{\rm m}$), converting it into a codeword bitstream (${ \bf s}_{\rm m}$), which is then fed back to the BS. The BS uses this feedback to reconstruct the downlink main channel (${ \bf h}_{\rm m}$).

\begin{figure*}[t]
    \centering
    \subfigure[Pilot design NN at the UE]{
        \includegraphics[width=0.45\linewidth]{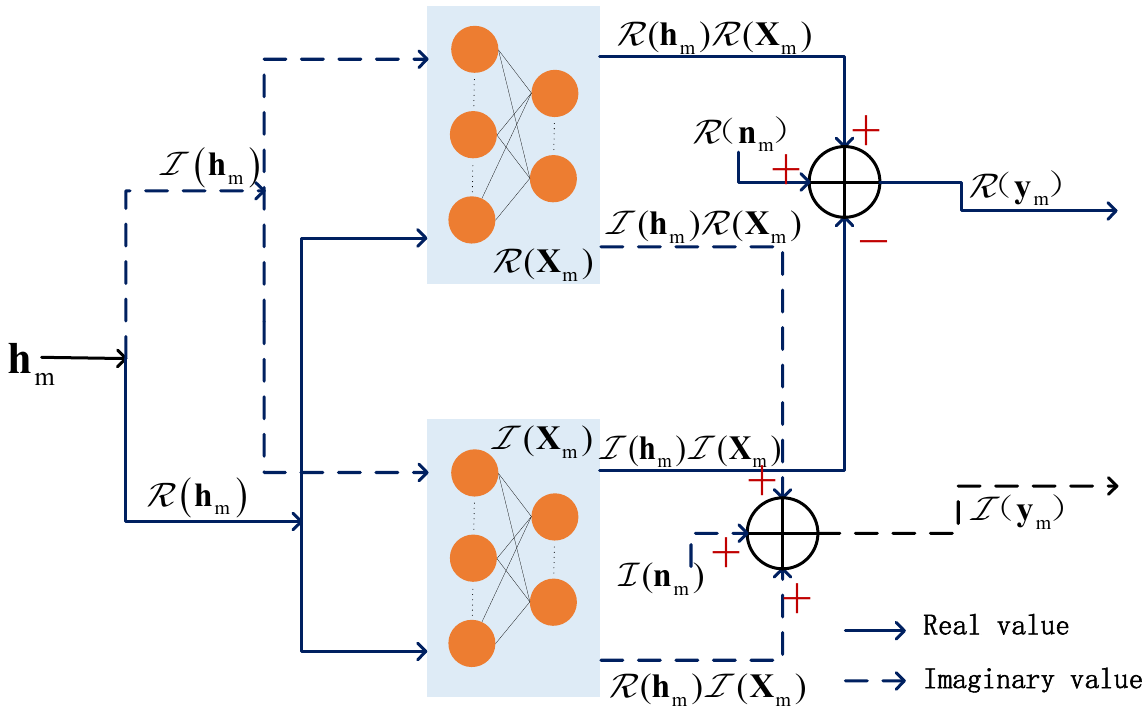}
        \label{Cpilot}
    }  \hfill
    \subfigure[Pilot signal compression NN at the UE]{
        \includegraphics[width=0.28\linewidth]{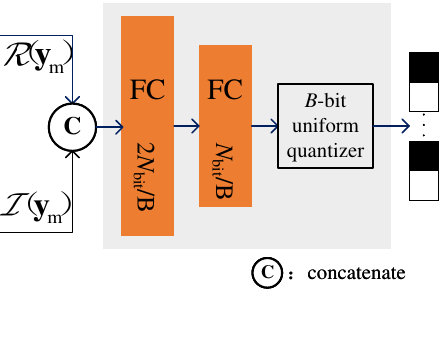}
        \label{UEcompressionNN}
    }  \hfill
    \subfigure[Channel reconstruction NN at the BS]{
        \includegraphics[width=0.21\linewidth]{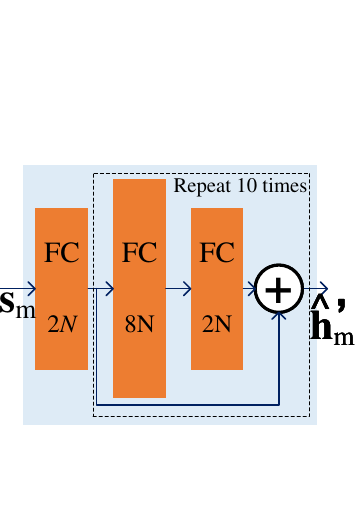}
        \label{BSrecNN}
    }
    \caption{Detailed NN architectures in the E2E-AI4CSI framework. (a): Illustration of the learning-based pilot design, which can be implemented using two linear FC layers without biases. (b): Illustration of the pilot signal compression module at the UE, consisting of two FC layers with $2N_{\rm bit}/B$ and $N_{\rm bit}/B$ neurons, along with a $B$-bit uniform quantizer. The activation functions applied to both FC layers are Sigmoid, ensuring the input is normalized within the range of $(0,1)$. (c): Illustration of the channel reconstruction module at the BS. The BS has significantly more computational power than the UE, allowing it to stack multiple FC layers to improve channel recovery.}
    \label{figure:E2EAI4CSI}
\end{figure*}

The process involves the following steps:
\begin{itemize}
    \item {\bf Pilot Generation at the BS:}
    From the perspective of compressive sensing, (\ref{pilotTrans}) represents a standard compression problem, as the pilot length $L_{\rm m}$ is much smaller than the number of antennas $N$ ($L_{\rm m} \ll N$). For real values, this compression can be achieved with a linear fully connected (FC) layer without biases, reducing the number of output neurons relative to the input. For complex values, the real and imaginary components are handled separately. Specifically, (\ref{pilotTrans}) is expressed as:
    \begin{align}
    \label{CpilotR} \hspace{-0.3cm}
    \mathcal{R}({\bf y}_{\rm m})  &=  \mathcal{R}({{\bf h}_{\rm m}}) \mathcal{R}({\bf X}_{\rm m}) -\mathcal{I}({{\bf h}_{\rm m}}) \mathcal{I}({\bf X}_{\rm m} )
    +\mathcal{R}({\bf z}_{\rm m}), \\
    \label{CpilotI} \hspace{-0.3cm}
    \mathcal{I}({\bf y}_{\rm m})  &=  \mathcal{I}({{\bf h}_{\rm m}}) \mathcal{R}({\bf X}_{\rm m}) +\mathcal{R}({{\bf h}_{\rm m}}) \mathcal{I}({\bf X}_{\rm m} )
    +\mathcal{I}({\bf z}_{\rm m}),
    \end{align}
    where $\mathcal{R}(\cdot)$ and $\mathcal{I}(\cdot)$ represent the real and imaginary components of a complex number, respectively.

    The operations in (\ref{CpilotR}) and (\ref{CpilotI}) can be implemented using two linear FC layers without biases \cite{9570376,8861085}, as shown in Figure \ref{Cpilot}. The inputs to these layers are the real and imaginary parts of the downlink channel (${\bf h}_{\rm m}$). The weight matrices of these FC layers correspond to the real and imaginary parts of the pilot symbol, i.e., $\mathcal{R}({\bf X}_{\rm m}) \in  \mathbb{R}^{N  \times L_{\rm m}} $ and $\mathcal{I}({\bf X}_{\rm m}) \in  \mathbb{R}^{N  \times L_{\rm m}}$.
    {During NN training, the weights of these layers are updated with training channel samples to minimize the training loss. Once trained, these weights form the pilot symbol ${\bf X}_{\rm m}$, which is then transmitted to the UE. The channels are no longer involved in the inference process whatsoever. }

    \item {\bf Pilot Signal Compression at the UE:}
    Due to limited computational power at the UE, a lightweight NN is employed to compress the received pilot signal ${\bf y}_{\rm m}$. The architecture of the pilot compression NN is illustrated in Figure \ref{UEcompressionNN}. Let $N_{\rm bit}$  represent the number of feedback bits and $B$ the quantization bit number. This module consists of two FC layers with $2N_{\rm bit}/B$ and $N_{\rm bit}/B$, followed by a $B$-bit uniform quantizer. Both FC layers use Sigmoid activation functions to normalize the input to the range $(0,1)$. The quantized codeword, ${\bf s}_{\rm m}$, is then transmitted back to the BS in bitstream form.

    \item {\bf Channel Reconstruction at the BS:}
    Since the BS has significantly more computational resources than the UE, it can stack multiple FC layers to improve the reconstruction of the main channel. As shown in Figure \ref{BSrecNN}, the quantized codeword ${\bf s}_{\rm m}$ is first passed through a linear FC layer with $2N$ neurons for initial channel recovery. The preliminary recovered channel is then processed by two FC layers with $8N$ and $2N$ neurons, respectively, using Tanh activation functions. Residual learning is employed, and this module is iterated 10 times to refine the reconstruction. The final output is the recovered main channel ${\widehat{\bf h}'_m}$.
    
\end{itemize}

{
The E2E-AI4CSI framework is trained using collected channel samples. As previously mentioned, this NN model consists of three key modules: pilot design at the BS, pilot signal compression at the UE, and channel reconstruction at the BS. These modules are trained simultaneously in an end-to-end manner to minimize the mean squared error (MSE) between the original and reconstructed channels.}

\section{PCEnet: Learning-based Position-domain Channel Extrapolation}
\label{PCEnetSection}
In this section, we first present the proposed primary  learning-based position-domain channel extrapolation framework, i.e., PCEnet. Following this, we address two practical challenges associated with PCEnet and introduce two novel techniques to tackle these challenges: one-sided real-time channel extrapolation and position label-free channel extrapolation.

 \begin{figure}
 \centering
\subfigure[Main motivation behind position-domain channel extrapolation using the channel-to-channel-characteristics mapping.] {
 \label{PositionBridge}
\includegraphics[width=0.78\linewidth]{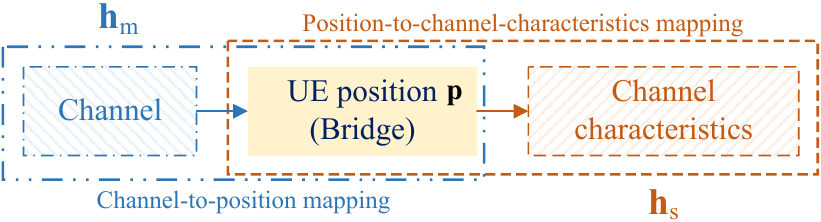}
}
\subfigure[Main workflow of position-domain channel extrapolation.] {
\label{MainWorkflow}
\includegraphics[width=0.8\linewidth]{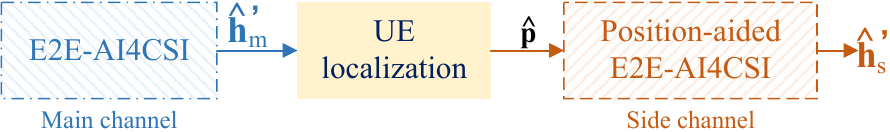}
}
\caption{{Main motivation and workflow of the proposed PCEnet, where the UE's position serves as the bridge connecting the main and side channels.}}
\label{figmotivationworkflow}
\end{figure}

 \begin{figure*}
 \centering
\subfigure[NN-based localization module.] {
 \label{BSlocNN}
\includegraphics[width=0.3\linewidth]{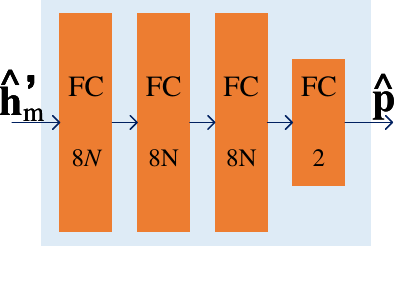}
}
\subfigure[{Proposed PCEnet framework.}] {
\label{PositionE2E}
\includegraphics[width=0.48\linewidth]{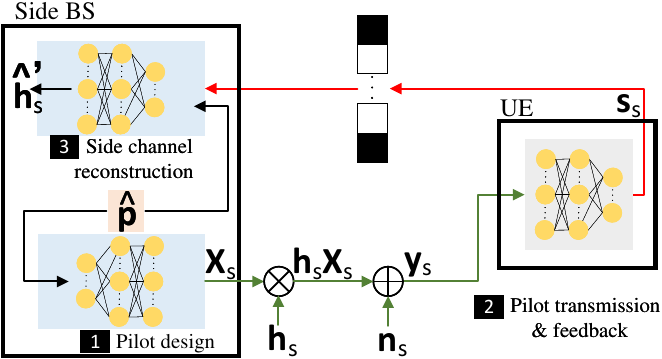}
}
\caption{(a): Illustration of the NN-based localization module. Three FC layers, each with $8N$ neurons, are employed to extract CSI features for position prediction, and an FC layer with two neurons is used to predict the UE's position. (b): Illustration of the proposed PCEnet framework for the side channel, where the estimated position ($\widehat{\bf p}$) is incorporated into the acquisition process of the side channel.}
\end{figure*}

\subsection{Primary PCEnet}

\subsubsection{Mapping Between Channel and Position}
\label{2mapping}

The wireless channel reflects the propagation environment and is shaped by its characteristics. In a stable environment, the channel is directly influenced by the UE's position. Consequently, some studies, such as \cite{9048929}, assume the existence of a position-to-channel mapping, meaning that the channel can be accurately inferred given the UE's position. However, this assumption does not hold in practical scenarios due to various unpredictable factors, such as atmospheric conditions and the presence of moving objects, which can affect signal propagation. Even for a stationary UE, channel conditions may fluctuate over time, rendering the position-to-channel mapping inaccurate.

\textbf{Channel-to-Position Mapping:}
While position alone is insufficient to accurately infer the channel, the reverse is true: the position can be inferred from the channel. In other words, the channel-to-position mapping, expressed as $ {\bf h}_{\rm m} \rightarrow {\bf p}$ and $ {\bf h}_{\rm s}\rightarrow {\bf p}$, is valid. This mapping can be seen as a CSI-based localization task \cite{6244790}, where the UE's position is inferred from the channel characteristics rather than the entire channel. For instance, in line-of-sight (LOS) scenarios, high-precision localization can be achieved using key channel features such as the angle and path loss of the LOS path, without relying on the complete channel information. 

\textbf{Position-to-Channel-Characteristics Mapping:} While the complete channel cannot be fully inferred from the UE's position alone, key channel characteristics can be derived from it. For example, the authors of \cite{10272348} propose a learning-based radio map concept, where a beamforming vector is directly generated from the UE's position using a pre-trained radio map constructed from a large dataset. Given that a mapping exists between the UE's position and partial channel characteristics, the authors of \cite{10597358} propose leveraging position data to enhance CSI feedback, thus reducing the feedback overhead caused by repeatedly transmitting channel information related to the UE's position. Moreover, the authors of \cite{9277535} incorporate position data into cross-band channel prediction to improve the quality of downlink CSI.

Although the signal propagation environments of the main and side channels differ, the UE's position remains constant in the scenario under consideration. This shared position can act as a bridge connecting the two mapping functions, thus reducing the overhead associated with CSI acquisition. The main motivation and framework of the proposed PCEnet are introduced in the following subsections.

\subsubsection{Main Motivation}

Given that the UE's position can be inferred from the channel and partial channel characteristics can be derived from the position, there exists a potential mapping from the channel to these partial characteristics. However, this channel-to-channel-characteristics mapping might initially seem redundant and of limited practical use, as channel characteristics are inherently part of the channel information. While this mapping may appear unnecessary in the context of a single channel, its utility becomes clear in multi-channel scenarios, such as in cell-free systems. In these cases, the mapping can significantly enhance the efficiency of CSI acquisition.

Figure \ref{PositionBridge} illustrates the core idea behind position-domain channel extrapolation using the channel-to-channel-characteristics mapping, where the UE's position acts as a bridge between the main and side channels, denoted as ${\bf h}_{\rm m}$ and ${\bf h}_{\rm s}$, respectively. On the one hand, the UE's position ($\bf p$) can be inferred from the main channel (${\bf h}_{\rm m}$), establishing the first mapping, known as the channel-to-position mapping. On the other hand, certain characteristics of the side channel (${\bf h}_{\rm s}$) can be derived from the corresponding UE's position ($\bf p$), representing the second mapping, referred to as the position-to-channel-characteristics mapping.

In the multi-channel scenarios under consideration, the main and side channels are typically uncorrelated due to the distinct propagation environments they encounter. However, the UE's position remains constant across both channels, making it a useful intermediary between the two. Assuming that the main channel (${\bf h}_{\rm m}$) has been obtained through the E2E-AI4CSI framework outlined in Section \ref{BenchmarkFrameworkSec}, the UE's position can be estimated based on the recovered main channel at the BS. This estimated position can then be used to generate the channel characteristics of the side channel (${\bf h}_{\rm s}$), facilitating its acquisition and reducing the overall acquisition overhead.

\subsubsection{Main Framework}
\label{MainFrameworkPosition0}
Figure \ref{MainWorkflow} illustrates the primary workflow of the proposed PCEnet. This process consists of three key steps: (1) main channel acquisition using the E2E-AI4CSI framework, (2) UE localization based on the recovered main channel (${\widehat{\bf h}'_m}$), and (3) side channel acquisition using the position-aided E2E-AI4CSI framework. Each of these steps will be discussed in detail in the following sections.

The first step is to obtain the main channel using the E2E-AI4CSI framework, as introduced in Section \ref{BenchmarkFrameworkSec}. To avoid redundancy, further elaboration on this step is omitted. Next, learning-based localization is employed to infer the UE's position from the recovered main channel ${\widehat{\bf h}'_m}$.
The NN architecture used for localization is shown in Figure \ref{BSlocNN}.
Three FC layers, each with  $8N$ neurons, are used to extract CSI features for position prediction. Since the UE height is fixed at 1.5 meters in this work, the UE's position, $\bf p$, is two-dimensional. Finally, an FC layer with two neurons is used to predict the UE's position, $\hat {\bf p}$, based on the extracted CSI features. All activation functions following the FC layers are ReLU.

{The predicted position $\hat {\bf p}$  is shared among BSs via a CPU  to assist in side channel acquisition.}
While a more advanced NN could enhance localization accuracy, this is not the primary focus of this study. { The objective is to demonstrate the feasibility of the proposed PCEnet framework. Thus, the simple NN architecture in Figure \ref{BSlocNN} is used for localization. For improved performance or more complex localization scenarios (e.g., urban environments), advanced localization NNs, such as those in \cite{9797028}, should be considered.
}

The main challenge lies in efficiently utilizing the estimated position ($\hat {\bf p}$) derived from the main channel to facilitate the acquisition of the side channel. As shown in Figure \ref{figmotivationworkflow}, the mapping from the UE's position to channel characteristics serves as the core concept of the position-aided E2E-AI4CSI framework. { This raises two key questions:
\begin{enumerate}
    \item[1)] What channel characteristics can be extracted from the UE's position, and how?
    \item[2)] How can these characteristics be leveraged to improve side channel acquisition?
\end{enumerate} }

Providing theoretical answers to these questions is challenging. However, NNs, trained end-to-end with large datasets, can automatically learn and utilize intrinsic relationships and data features, bypassing the need for manual feature extraction. Although NNs are sometimes perceived as ``black-box'' models, they offer significant potential for solving tasks that are difficult to model analytically \cite{10639525,9277535}. Consequently, in the proposed position-aided E2E-AI4CSI framework, the estimated position ($\widehat{\bf p}$) is directly fed into the NNs, which autonomously learn and utilize the channel characteristics associated with the position data, thereby addressing the aforementioned challenges.

Figure \ref{PositionE2E} shows the proposed PCEnet framework for the side channel using the position-aided E2E-AI4CSI framework. The three main steps are as follows:
\begin{itemize}
\item {\bf Pilot Design:} A pilot design NN at the side BS generates the pilot symbol (${\bf X}_{\rm s}$) based on the estimated position ($\widehat{\bf p}$) from the main channel. Assuming the pilot length of the side channel is $L_{\rm s}$, the pilot design NN first uses an FC layer with $4L_{\rm s}N$ neurons to extract features from the estimated position. Then, two parallel FC layers, each with $L_{\rm s}N$ neurons, are used to generate the real and imaginary components of the pilot symbol (${\bf X}_{\rm s}$),  respectively. A normalization layer ensures that the pilot symbol adheres to power constraints, and all FC layers are followed by the Tanh activation function.

\item {\bf Pilot Transmission and Feedback:}
Position-dependent pilot symbols are transmitted from the side BS to the UE. The UE then compresses and feeds back the received pilot signal to the side BS using the same NN architecture employed in the E2E-AI4CSI framework.

\item {\bf Side Channel Reconstruction:}
Upon receiving the feedback bitstream, the side BS reconstructs the side channel by leveraging both the feedback information and the UE's position information. As noted in \cite{9277535,10597358}, position and channel information are distinct modalities, and fusing such multi-modal information is a complex task. Inspired by \cite{10597358}, we adopt a hybrid fusion strategy---a flexible multi-modality fusion technique that integrates feature-level and data-level fusion. Specifically, two FC layers with $4N$ and $N$ neurons, followed by a Sigmoid activation function, are used to generate features from the estimated position. These features are then concatenated with the feedback information and fed into the reconstruction NN, which is identical to the one used in the E2E-AI4CSI framework.

\end{itemize}

\subsubsection{Training Strategy}
\label{TrainingStrategy1}

The proposed PCEnet framework consists of three key steps, as shown in Figure \ref{MainWorkflow} and described earlier. This sequential three-step design is adopted to effectively capture and utilize the two mappings between the channel and position, as introduced in Section \ref{2mapping}. By training separate NNs for each step, the entire framework is guided to efficiently acquire the side channel, as illustrated in Figure \ref{figmotivationworkflow}. 

{
First, the E2E-AI4CSI framework, used for main channel acquisition, is trained on collected main channel samples. This NN model consists of three modules: pilot design at the BS, pilot signal compression at the UE, and channel reconstruction at the BS. These modules are trained simultaneously in an end-to-end manner to minimize the MSE between the original main channel ${{\bf h}_m}$ and the reconstructed main channel ${\widehat{\bf h}'_m}$.}

Next, the localization NN, illustrated in Figure \ref{BSlocNN}, is trained using pairs of recovered main channel data and their corresponding position labels, i.e., $\{{\widehat{\bf h}'_m}, {\bf p} \}$. {The} goal of this training process is to minimize the MSE between the ground truth position ($\bf p$) and the predicted position ($\widehat{\bf p}$).

Finally, the position-aided E2E-AI4CSI NN, used for side channel acquisition, is trained using the collected side channel data (${{\bf h}_s}$) and the position estimate ($\widehat{\bf p}$) derived from the recovered main channel ${\widehat{\bf h}'_m}$. As in the first training step, the three NN modules within this framework are trained together in an end-to-end fashion, with the objective of minimizing the MSE between the original side channel and the reconstructed side channel.

\begin{algorithm}[t]
\caption{\label{algorh1}{Inference Process of the Proposed PCEnet}}
\begin{algorithmic}[1]
    \Statex \hrulefill \textbf{Stage 1: Main Channel Acquisition} \hrulefill
    \State \hskip1em \textit{Main BS}: Transmit the learned pilot symbol $ \mathbf{X}_{\rm m} $ to the UE
    \State \hskip1em \textit{UE}: Compress and feed back the received signal $ \mathbf{y}_{\rm m} $
    \State \hskip1em \textit{Main BS}: Estimate main channel $ \mathbf{h}_{\rm m} $ from received codeword $ \mathbf{s}_{\rm m} $

    \Statex \hrulefill \textbf{Stage 2: Side Channel Acquisition} \hrulefill
    \State \hskip1em \textit{Main BS}: Estimate UE position $ \mathbf{p} $ from $ \hat{\mathbf{h}}_{\rm m}^{'} $
    \State \hskip1em \textit{Main BS}: Send the estimated position $ \hat{\mathbf{p}} $ to the side BS
    \State \hskip1em \textit{Side BS}: Design pilot $ \mathbf{X}_{\rm s} $ based on the received $ \hat{\mathbf{p}} $
    \State \hskip1em \textit{Side BS}: Transmit the designed pilot $ \mathbf{X}_{\rm s} $ to the UE
    \State \hskip1em \textit{UE}: Compress and feed back received pilot signal $ \mathbf{y}_{\rm s} $
    \State \hskip1em \textit{Side BS}: Estimate side channel $ \mathbf{h}_{\rm s} $ from received codeword $ \mathbf{s}_{\rm s} $
\end{algorithmic}
\end{algorithm}

The inference process (deployment) of the proposed PCEnet is shown in {\bf Algorithm \ref{algorh1}}. In the main channel acquisition, the main BS transmits the learned pilot symbol $ \mathbf{X}_{\rm m} $ to the UE. The UE then compresses and feeds back the received pilot signal $ \mathbf{y}_{\rm m} $. Then, the main BS estimates the main channel $ \mathbf{h}_{\rm m} $ based on the received codeword $ \mathbf{s}_{\rm m} $. In the side channel acquisition, the main BS estimates the UE position $ \mathbf{p} $ using the previously recovered main channel $ \hat{\mathbf{h}}_{\rm m}^{'} $, and subsequently shares this estimated position $ \hat{\mathbf{p}} $ with the side BS through a CPU. The side BS designs a pilot $ \mathbf{X}_{\rm s} $ based on the received $ \hat{\mathbf{p}} $ and transmits it to the UE. The UE compresses and feeds back the received pilot signal $ \mathbf{y}_{\rm s} $, enabling the main BS to estimate the side channel $ \mathbf{h}_{\rm s} $ using the corresponding codeword $ \mathbf{s}_{\rm s} $.

\subsubsection{{Robustness}}

The proposed framework establishes a universal approach that efficiently links the main channel to side channel characteristics across various propagation environments. However, due to its data-driven nature, the NN-based localization component is highly dependent on the mapping between the user position and channel characteristics, which can be significantly affected by changes in the propagation environment. As a result, NNs trained within this framework may struggle with substantial environmental variations.
To enhance robustness against such changes, online training can be employed. However, to expedite this process and minimize computational and data collection overhead, advanced techniques such as meta-learning and data augmentation should be explored.

With the mapping between the main channel and side channel characteristics established, the proposed framework can be adapted to various antenna network topologies. For instance, in reconfigurable intelligent surface (RIS)-assisted systems \cite{10801257}, where RIS improves capacity, the network structure differs from the cell-free system studied in this work. In such a scenario, the direct BS-UE link serves as the main channel, while the RIS-UE link functions as the side channel. Here, the UE's position bridges the BS-UE and RIS-UE channels, allowing the proposed position-domain channel extrapolation to be seamlessly applied. This demonstrates the versatility and resilience of the proposed framework across different antenna network topologies. However, adapting the framework to new topologies may require modifications to the NN architecture to align with the specific characteristics of the new antenna configurations.

\subsection{One-Sided Real-Time PCEnet}

In the proposed PCEnet framework above, the acquisition of the main and side channels is performed sequentially. The side channel acquisition depends on the position estimation obtained from the main channel. Simultaneously, the estimated position is integrated into the entire side channel acquisition process, including both pilot design and channel reconstruction. This approach, referred to as a two-sided method, has the potential to maximize the benefits of position-based channel extrapolation. However, this process can be complex and may not be feasible in simpler systems.

The primary challenge is how to effectively integrate the estimated position into the side channel acquisition process without relying on the more complex two-sided approach. The estimated position can be treated as side information that correlates with the side channel to be acquired. Drawing inspiration from the concept of deep image compression with side information at the decoder \cite{ayzik2020deep}, we propose incorporating the estimated position solely within the side channel reconstruction module at the BS.

In this approach, both the pilot design and pilot signal compression remain unchanged from the E2E-AI4CSI framework. Specifically, the pilot symbol is fixed and generated by FC layers, as shown in Figure \ref{Cpilot}. Since the estimated position is only integrated into the side channel recovery module---and the pilot symbol does not require prior position estimation from the main channel---we refer to this approach as one-sided real-time PCEnet.

\subsection{Position Label-Free PCEnet}

\subsubsection{Motivation}

In the proposed PCEnet framework, the position estimated from the main channel serves as a crucial bridge between the main and side channels. During the supervised training of the localization NN, accurate position labels are essential. However, collecting such labels can be time-consuming and impractical in many scenarios, making the previously proposed PCEnet approach infeasible in some cases.

Recently, a novel localization technique called channel charting has been introduced \cite{8444621,10155724}. Unlike traditional learning-based localization methods, which aim to accurately determine the UE's position and require labeled training data, channel charting focuses on deriving a pseudo-position for the UE through CSI dimensionality reduction, without needing position labels. To achieve this, unsupervised dimensionality reduction algorithms, such as autoencoders or principal component analysis (PCA), are used to map high-dimensional CSI to a lower-dimensional space. In this space, the position of each point represents the relative position of the UE.

In the scenario under consideration, the absolute position of the UE is not necessary; rather, its relative position is sufficient for our purposes. Thus, the problem shifts from determining the UE's absolute position to identifying its relative position. As previously mentioned, dimensionality reduction techniques like autoencoders can be employed to derive this relative position. In such a setup, an encoder compresses the original CSI into a lower-dimensional representation, while a decoder reconstructs the CSI from this representation by minimizing the MSE between the original and reconstructed CSI. However, a standard autoencoder may not ensure that the low-dimensional code (the encoder's output) corresponds accurately to the UE's position space, as it lacks a regularizer or prior information to guide the encoding process.
{Therefore, the unique structural characteristics of the considered scenario should be leveraged to address this challenge.}

\subsubsection{Proposed Method}
{
Inspired by channel charting and considering the presence of two distinct channel types---main channel ${\bf h}_{\rm m}$ and side channel ${\bf h}_{\rm s}$---we propose a novel unsupervised channel charting approach to derive relative positions without requiring labeled data. This method leverages unsupervised learning to infer relative positions, enabling a label-free approach to bridge the main and side channels within the PCEnet framework. }

\begin{figure}[t]
    \centering
    \includegraphics[width=0.99\linewidth]{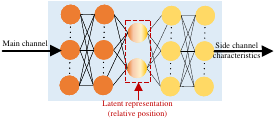}
    \caption{Illustration of the mapping from main-channel characteristics to side-channel characteristics using two NNs. The latent representation (${\bf{s}}_{\rm p}$), which is the output of the first NN, is strongly associated with the UE's position.}
    \label{ChannelChart}
\end{figure}

{
As previously mentioned, directly inferring the side channel from the main channel is infeasible. However, as discussed in Section \ref{2mapping}, a mapping exists between the main-channel characteristics and side-channel characteristics, mediated by the UE's position. 
{This process can be formulated as a specialized autoencoder within channel charting.}
To establish this mapping, we propose using two NNs, as illustrated in Figure \ref{ChannelChart}. The first NN (encoder in channel charting) estimates the absolute or relative position of the UE from the main channel. The second NN (decoder in channel charting) utilizes this estimated position to predict side-channel characteristics. Since main and side channels propagate in uncorrelated environments, the UE's position is the only common link between them. Training these two NNs together ensures that the latent representation (output of the first NN) is strongly correlated with the UE's position, facilitating reliable side-channel prediction.}
 
Another key aspect is defining the side channel characteristics. If the UE's position is known, characteristics such as signal direction and path loss---considered channel characteristics---can be inferred. These properties can be represented by the channel magnitude in the angular domain. Therefore, in this study, we choose the side channel magnitude in the angular domain \cite{9570376} as the target characteristic. 

Thus, the output of the NNs in Figure \ref{ChannelChart} corresponds to the side channel magnitude in the angular domain. The detailed architecture of the NNs is as follows: 
\begin{itemize}
    \item The first NN consists of three FC layers with $4N$ , each with $4N$ neurons, followed by ReLU activation functions, to extract localization features from the main channel. A final FC layer with two neurons, followed by a Sigmoid activation function, compresses the CSI into a 2-dimensional latent representation, ${\bf{s}}_{\rm p}$, which is expected to reflect the UE's relative position through end-to-end training.
    
    \item The second NN reconstructs the side channel characteristics---specifically, the side channel magnitude in the angular domain---based on the 2-dimensional latent representation. The architecture of this NN mirrors the reconstruction module used in the E2E-AI4CSI framework (Figure \ref{BSrecNN}), but with half the number of neurons in the final layer, as only the side channel magnitude is being recovered. 
\end{itemize}

\subsubsection{Training and inference strategies}

The training strategy for the proposed position label-free channel extrapolation involves three sequential steps, similar to the approach outlined in Section \ref{TrainingStrategy1}. The detailed NN training strategy is as follows: 
\begin{itemize}
    \item {\bf Training an NN to Acquire Main Channel:} The first step involves training the E2E-AI4CSI NN for main channel acquisition. This training process follows the methodology described in Section \ref{TrainingStrategy1}. 
        
    \item {\bf Training an NN for Relative Position Estimation:} To obtain the relative position from the main channel, the NN shown in Figure \ref{ChannelChart} is trained end-to-end to map the main channel to the side channel characteristics. The input to this NN is the output from the E2E-AI4CSI NN trained in the previous step, i.e., ${\widehat{\bf h}'_m}$, and the output is the perfect side channel magnitude in the angular domain.

    \item {\bf Training an NN for Side Channel Acquisition:} Finally, the 2-dimensional latent representation (${\bf{s}}_{\rm p}$) generated by the first NN is combined with the side channel data (${\bf h}_{\rm s}$) to train the position-aided E2E-AI4CSI NN for side channel acquisition. 
\end{itemize}

During inference, the position label-free channel extrapolation follows a process similar to that of the original PCEnet, as shown in Figure \ref{PositionE2E}. The main difference lies in the type of position information and acquisition method used. In the original PCEnet framework, the absolute position estimated by the localization NN (Figure \ref{BSlocNN}) is used for extrapolation. In contrast, the position label-free channel extrapolation uses the relative position inferred by the first NN (left side of Figure \ref{ChannelChart}) for the same purpose.

\section{Numerical Simulation Results and Discussion}
\label{s4}
This section presents an overview of the simulation settings, including the generation of CSI data and the details of NN training. Following this, the performance of the proposed learning-based channel extrapolation frameworks is evaluated and discussed.

\subsection{Simulation Settings}

\subsubsection{Channel Generation}

\begin{figure}[t]
    \centering
    \includegraphics[width=0.95\linewidth]{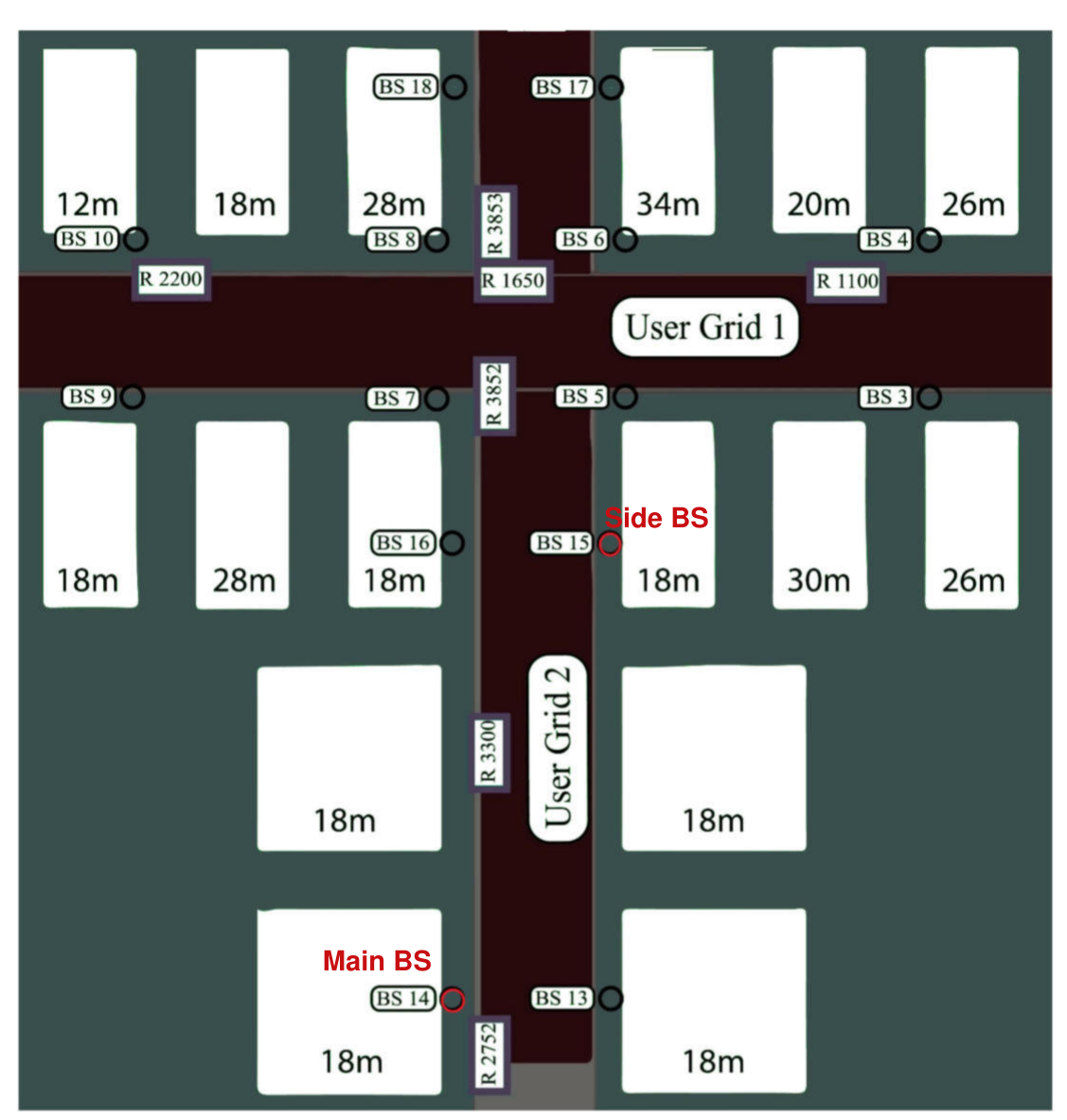}
    \caption{Top view of the ``O1'' scenario from \cite{alkhateeb2019deepmimo}. The 14th and 15th BSs are selected as the main and side BSs, respectively. The UEs are located within the second user grid, spanning from Row 3,000 to Row 3,800.}
    \label{O1topV2}
\end{figure}

In this study, the position of the UE is crucial for establishing the relationship between the main and side channel characteristics. As a result, both the main and side channels, along with the corresponding UE positions, are required, making commonly used public CSI datasets (e.g., \cite{wirelessintelligence}) unsuitable for our purposes. To address this, we utilize the publicly available DeepMIMO framework \cite{alkhateeb2019deepmimo}, which is based on Wireless InSite ray tracing software, to generate simulated channels and corresponding positional data. An outdoor ray tracing scenario, referred to as ``O1'', is selected. This scenario includes a total of 18 BSs. For the simulations, we specifically select the 14th and 15th BSs as the main and side BSs, respectively, as illustrated in Figure \ref{O1topV2}.
The UEs are distributed within the second user grid, spanning from Row 3,000 to Row 3,800, with each row containing 181 UEs spaced 20 cm apart. This results in a total of $801 \times 181 = 144,981$ UEs.

{
It is important to note that in this study, one BS is randomly assigned as the main BS, while the remaining BSs are designated as side BSs; BS selection is not explicitly considered. However, the selection of the main BS should be based on the gain achieved through the proposed position-domain channel extrapolation. Since UE position serves as the bridge connecting main and side channel characteristics, localization accuracy significantly impacts the framework’s performance. Achieving high-accuracy localization in complex environments remains a challenge. Therefore, a BS with a LOS path to the UE is preferred as the main BS, as it provides more reliable localization information. Additionally, the proposed framework significantly reduces the acquisition overhead of the side channel by leveraging prior knowledge of its characteristics. Thus, in scenarios where BSs have different numbers of antennas, BSs with more antennas are preferable as side BSs, further minimizing CSI acquisition overhead across the entire cell-free system. A more in-depth study of BS selection strategies is reserved for future research. }

For the simulations, downlink transmission is conducted at a frequency of 3.5 GHz, considering a total of ten propagation paths. The BSs are positioned at a height of 6 m, while the UEs are positioned at 2 m. Additionally, the BSs are equipped with 32 ULA transmitting antennas, spaced at half-wavelength intervals.
A total of 144,981 downlink CSI samples and their corresponding UE positions are generated for the ``O1'' scenario using the DeepMIMO framework. The generated dataset is divided into training, validation, and test sets, consisting of 85\%, 5\%, and 10\% of the samples, respectively.

 \begin{figure}[t]
    \centering
    \includegraphics[width=0.98\linewidth]{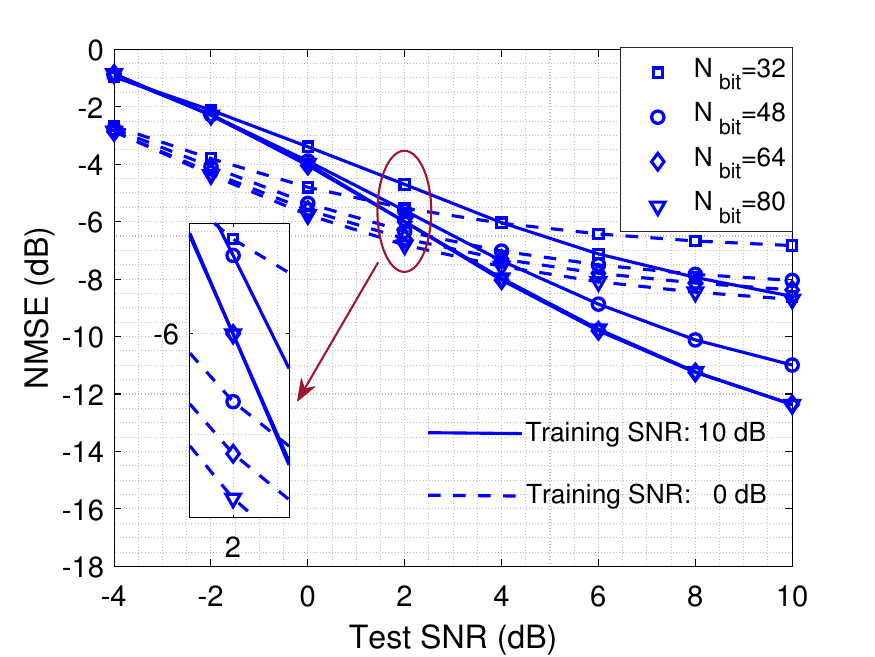}
    \caption{NMSE (dB) performance of the E2E-AI4CSI framework with different feedback bit number ($N_{\rm bit}$) across various channel estimation SNRs. The pilot length $L_{\rm m}$ is set as 8.}
    \label{BenchmarkPlot0}
\end{figure}

{
The distribution of simulated data does not perfectly match that of practical systems, leading to potential performance degradation due to generalization issues. To address this, a two-step training process, comprising offline pre-training and online training, is required. As an example, we consider the digital twin-based NN training approach from \cite{10622316}.
First, a digital twin replica of the target practical scenario is constructed on the infrastructure side. This replica is used to generate training channel data via ray tracing techniques, which train the AI-based CSI acquisition model.
During deployment, UEs estimate the downlink CSI and feed back high-quality CSI to the BS, albeit at the cost of increased overhead.\footnote{Some studies, such as \cite{9714227}, propose training CSI acquisition NNs using uplink CSI to eliminate the need for feedback during data collection.} The BS collects CSI matrices from the UEs, and once a sufficient number of full CSI matrices have been received, it constructs a refinement dataset to fine-tune the pre-trained NN model, improving its generalization to real-world propagation environments. 
}
\subsubsection{NN Training Details}
All simulations are carried out on an NVIDIA DGX-2 workstation using the TensorFlow 2.13 library. A four-bit uniform quantization (i.e., $B=4$) is applied during the feedback process, with the gradient of the quantization set to one to enable backpropagation during the quantization operation. At the beginning of training, all NN parameters are initialized using the Glorot uniform initializer. The Adam optimization algorithm is employed for updating the NN parameters during training.

For all NN training in this work, the learning rate is set to 0.001, and the batch size is set to 512. The E2E-AI4CSI framework (Figure \ref{E2Ebenchmark}) for main channel acquisition and the proposed PCEnet framework (Figure \ref{PositionE2E}) for side channel acquisition are both trained for 500 epochs. In contrast, the supervised localization NN (Figure \ref{BSlocNN}) is trained for 300 epochs, and the NNs responsible for mapping main-channel-to-side-channel characteristics (Figure \ref{ChannelChart}) are trained for 100 epochs.

The NN model that performs best on the validation dataset is saved as the final trained model. All NN training processes in this study utilize the MSE loss function. Consistent with commonly adopted practices in related works \cite{9931713}, the normalized MSE (NMSE) is used to evaluate the accuracy of CSI acquisition. For example, for the main channel, the NMSE is defined as follows:
\begin{equation}
\mathrm{NMSE}=\frac{1}{S} \sum_{i=1}^{S} \frac{\|{\bf h}_{\rm m}^{(i)}-\widehat{{\bf h}}_{\rm m}'^{(i)}\|_{2}^{2}}{\|{\bf h}_{\rm m}^{(i)}\|_{2}^{2}} ,
\end{equation}
where ${\bf h}_{\rm m}^{(i)}$ and $\widehat{{\bf h}}_{\rm m}'^{(i)}$  denote the $i$-th original and estimated main channels, respectively, $S$ is the number of samples, and $\|\cdot\|_2$ represents the Euclidean norm.

\subsection{Performance Evaluation of the Primary PCEnet Framework}

This subsection begins with a presentation and evaluation of the E2E-AI4CSI framework. Following this, we introduce and analyze the performance of our proposed primary PCEnet method.

 \begin{figure}[t]
    \centering
    \includegraphics[width=0.98\linewidth]{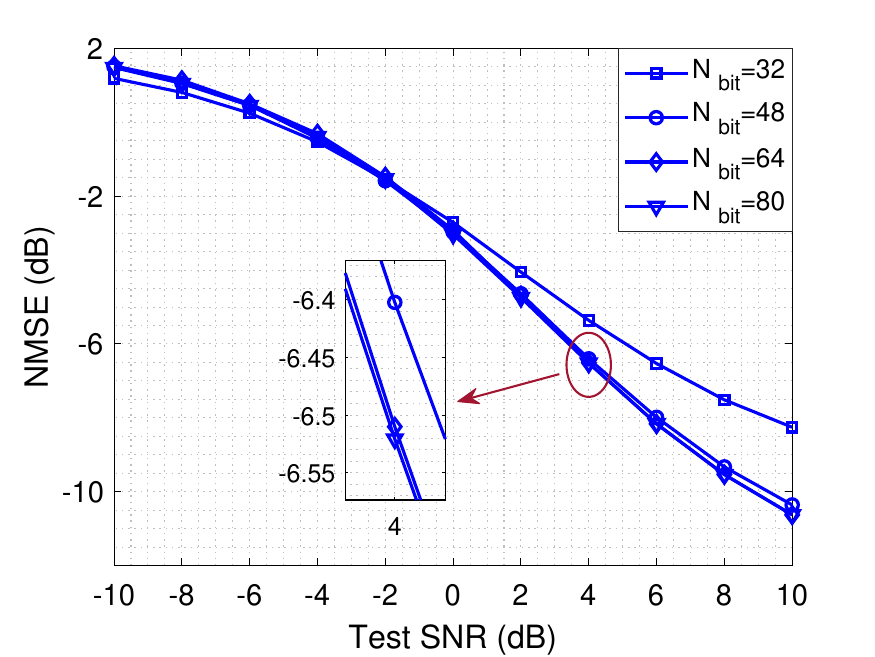}
    \caption{{NMSE (dB) performance of the E2E-AI4CSI framework with different feedback bit number ($N_{\rm bit}$) across various channel estimation SNRs. The training SNR is 10 dB, and the pilot length $L_{\rm m}$ is set as 6.}}
    \label{Benchmark0_pilot}
\end{figure}

\subsubsection{Performance of the E2E-AI4CSI Framework}
In this part, we evaluate the benchmark E2E-AI4CSI framework, using the main channel acquisition as an example. As shown in Figure \ref{BenchmarkPlot0}, we examine the NMSE performance of the E2E-AI4CSI framework with varying numbers of feedback bits ($N_{\rm bit}$) across different channel estimation signal-to-noise ratios (SNRs). The pilot length $L_{\rm m}$ is set to 8. Solid lines represent training at 10 dB SNR, while dashed lines represent training at 0 dB SNR.

For a given pilot length, feedback accuracy improves as the number of feedback bits increases, but eventually reaches a plateau. For example, at a 10 dB training and testing SNR, the NMSE values for 64 and 80 feedback bits are $-12.36$ dB and $-12.37$ dB, respectively. Similar trends are observed when the training SNR is lowered to 0 dB.
{To further validate this observation, we examined a scenario with a different pilot length of 6. The simulation results, depicted in Figure \ref{Benchmark0_pilot},  confirm a similar trend.} The observed plateau occurs because increasing the number of feedback bits beyond a certain threshold does not significantly enhance accuracy, as the limited information in the received pilot signal constrains further improvements. The feedback accuracy plateau is observed at 64 feedback bits, corresponding to $ 2\times B \times L_{\rm m} = 2\times 4 \times 8$. The received pilot signal undergoes pre-processing with several FC layers\footnote{We also simulated direct quantization and feedback of the received signal, but this led to a significant decrease in accuracy. Pre-processing with FC layers before quantization helps mitigate the negative impact of channel estimation and quantization noise.} before being quantized to 4 bits and sent back to the BS. This aligns the pilot and feedback overheads, so for the following evaluations, we focus on the case where $N_{\rm bit} = 2 B L_{\rm m}$.

\begin{figure}[t]
    \centering
    \includegraphics[width=0.98\linewidth]{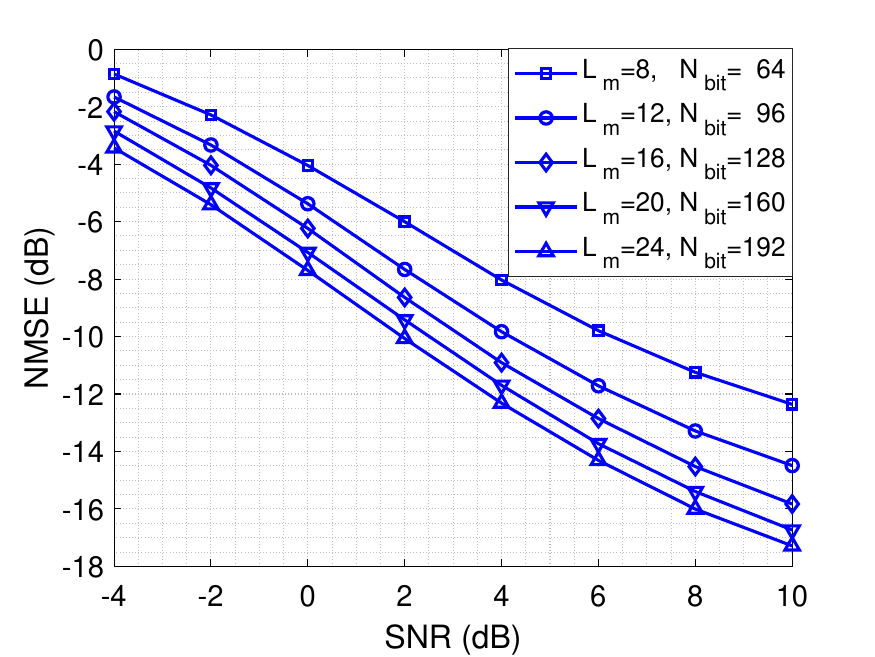}
    \caption{NMSE (dB) performance of the E2E-AI4CSI framework with different pilot lengths ($L_{\rm m}$) versus channel estimation SNRs, where $N_{\rm bit} = 2BL_{\rm m}$ and the training SNR is set to 10 dB.}
    \label{BenchmarkPlot}
\end{figure}

Additionally, at higher test SNRs (e.g., 6 dB), the NN trained with a 10 dB SNR outperforms the one trained with 0 dB SNR. Conversely, at lower test SNRs (e.g., 0 dB), the NN trained at 0 dB shows better performance. While training an NN for each channel estimation SNR would be optimal, it is impractical in real-world systems due to the unknown SNR and complexity. {A potential solution is to train the model over a range of SNRs to enhance robustness. However, given the objectives of this study, we adopt a fixed training SNR of 10 dB for the subsequent evaluations.}

\begin{figure}[t]
    \centering
    \includegraphics[width=0.98\linewidth]{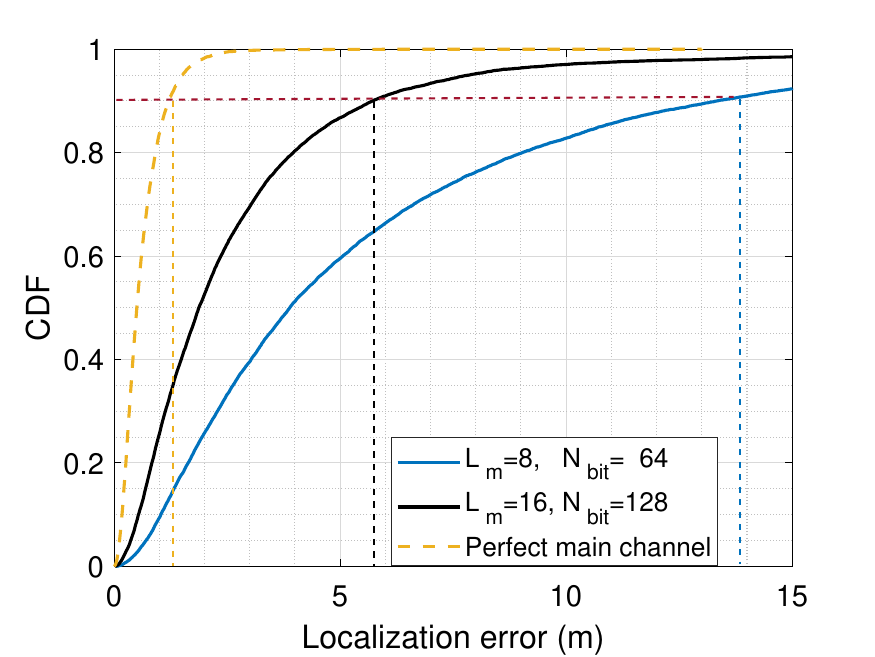}
    \caption{CDF illustration of the main channel-based localization. The training and the test SNRs for the main channel acquisition are both 10 dB. The accuracy of localization is heavily influenced by the quality of the main channel.}
    \label{LocCDF}
\end{figure}

As shown in Figure \ref{BenchmarkPlot}, we also evaluate the NMSE performance of the E2E-AI4CSI framework for various pilot lengths ($L_{\rm m}$) versus channel estimation SNRs. As expected, increasing the pilot length improves main channel accuracy, though the magnitude of improvement diminishes as the pilot length increases further.

\subsubsection{Performance of PCEnet}

As mentioned in Section \ref{MainFrameworkPosition0}, the proposed PCEnet framework consists of three steps. In the first step, we obtain the main channel using the E2E-AI4CSI framework, the results of which were discussed earlier. The second and third steps involve estimating the UE's position based on the obtained main channel and utilizing that position to aid in acquiring the side channel. We will now present and analyze the simulation results of these last two steps.

Figure \ref{LocCDF} shows the cumulative distribution function (CDF) for the second step---main channel-based localization. Solid lines represent localization using the main channel obtained from the E2E-AI4CSI NN with different overheads, while dashed lines represent localization based on the perfect main channel. Both the training and test SNRs for the main channel acquisition are 10 dB. The figure illustrates that localization accuracy is heavily influenced by the quality of the main channel. If the perfect main channel is available, the mean localization error is 0.62 m, with 90\% of the samples having an error below 1.5 m. However, when the pilot length and feedback bit number are 16 and 128, the mean error increases to 2.84 m, with 90\% of errors below 5.8 m. Reducing the pilot length to 8 and feedback bits to 64 results in a mean error of 6.09 m, with 90\% of errors below 13.9 m.

These results highlight the critical importance of main channel quality. Poor main channel accuracy limits localization performance, which in turn affects side channel acquisition. Nevertheless, simulations presented in \cite[Table VIII]{10597358} indicate that even with a mean position error of up to 2.5 m, position information can still assist with channel feedback. In this study, we set the pilot length and feedback bit number for the main channel to 16 and 128, respectively, for subsequent evaluations. Moreover, future research may explore improving localization performance without increasing acquisition overhead by jointly considering main channel acquisition and UE localization as discussed in \cite{10597358}.

Figure \ref{positionAid} shows the NMSE performance of the proposed PCEnet framework for the side channel. The pilot length for the side channel, $L_{\rm s}$, is set to 4 during the evaluation of the proposed method.
{
In addition to comparing against the E2E-AI4CSI framework presented in Section \ref{BenchmarkFrameworkSec}, we also evaluate PCEnet against a conventional non-learning-based method. Specifically, in this baseline approach, the main and side channels are estimated at the UE using a centralized minimum mean squared error (MMSE) algorithm and then transmitted directly to the BSs without compression, assuming perfect feedback. For this comparison, the pilot lengths for both the main and side channels, denoted as $L_{\rm m}$ and $L_{\rm s}$, are set to 16.} Key observations include:

\begin{figure}[t]
    \centering
\includegraphics[width=0.98\linewidth]{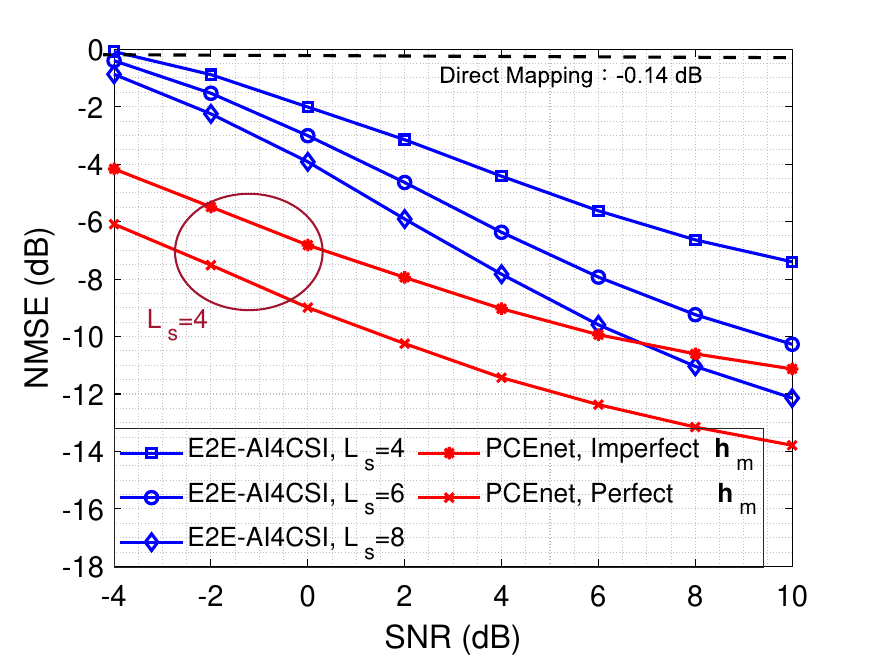}
    \caption{{NMSE (dB) performance of the proposed PCEnet framework for the side channel. The proposed method reduces the channel acquisition overhead (including pilot and feedback) by half without sacrificing performance.}}
    \label{positionAid}
\end{figure}

\begin{figure}[t]
    \centering
    \includegraphics[width=0.98\linewidth]{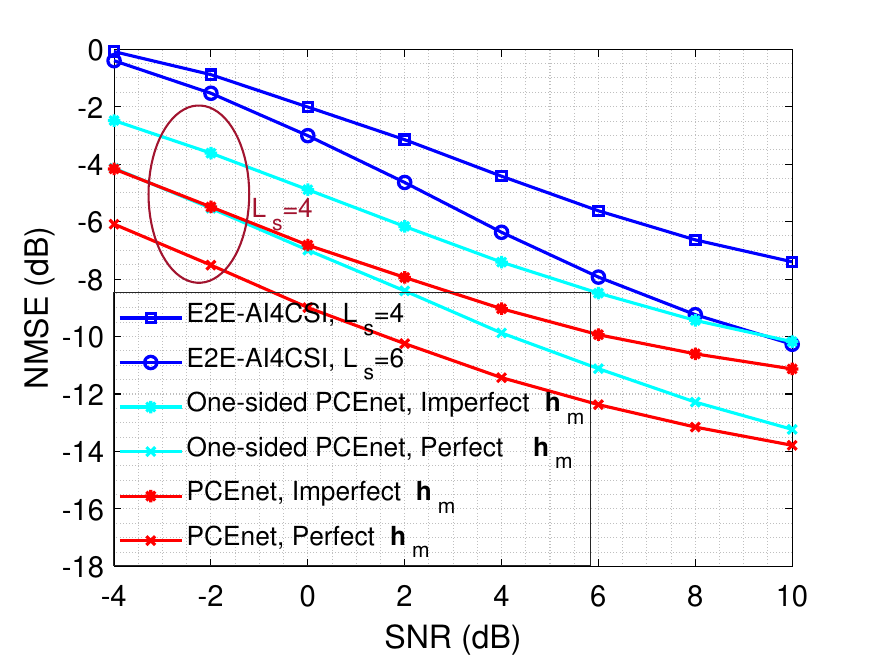}
    \caption{NMSE (dB) performance of the proposed one-sided real-time PCEnet for the side channel.}
    \label{oneSided}
\end{figure}

\begin{figure*}[t]
    \centering
    \subfigure[Ground truth]{
        \includegraphics[width=0.3\linewidth]{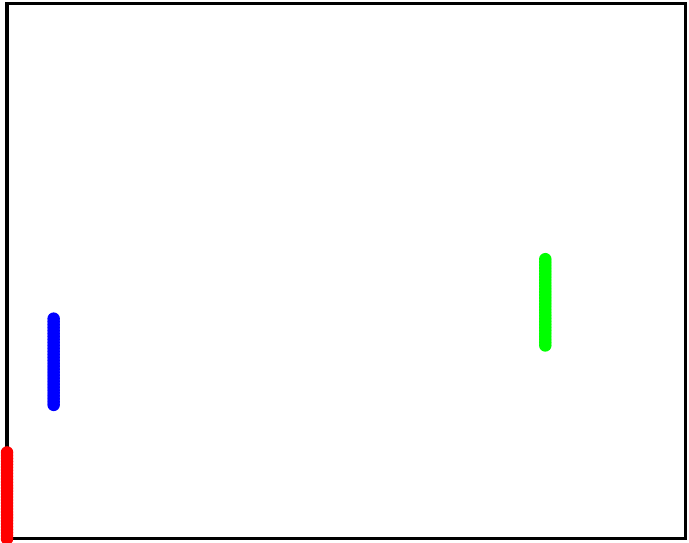}
        \label{OriginalLOC}
    }
    \subfigure[Vanilla autoencoder]{
        \includegraphics[width=0.3\linewidth]{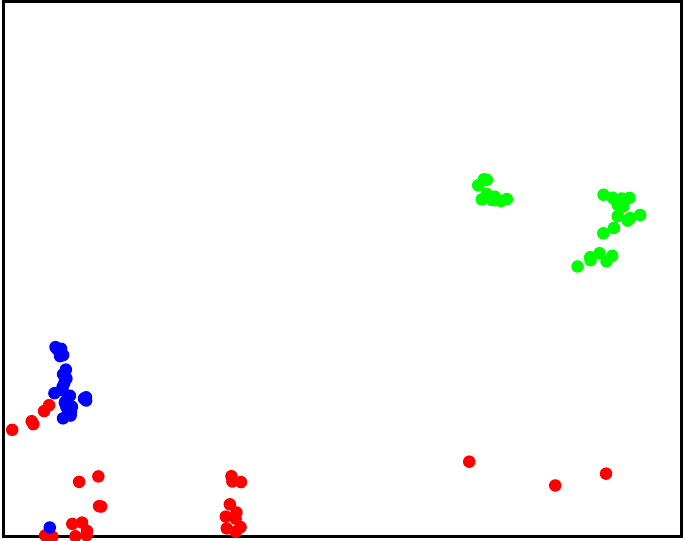}
        \label{CCae0}
    }
    \subfigure[Proposed]{
        \includegraphics[width=0.3\linewidth]{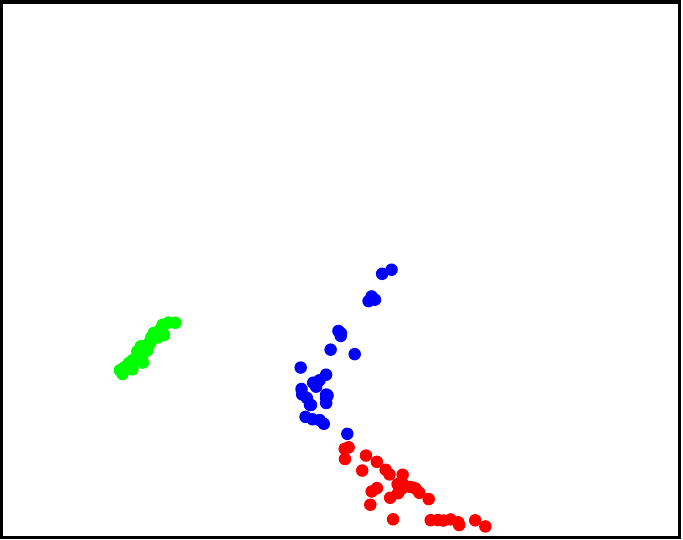}
        \label{CCae}
    }
    \caption{Illustration of the UE's ground truth position and the estimated relative position. (a): UE's ground truth position; (b): relative UE's position generated by the vanilla autoencoder; (c): relative UE's position generated by the innovative main-channel-to-side-channel characteristics mapping. By leveraging main-channel-to-side-channel characteristics mapping, UEs within the same group are positioned closely together, effectively demonstrating relative position relationships.}
    \label{CCresult}
\end{figure*}

\begin{itemize}
    \item {{\bf Significant Improvement in Side Channel Acquisition compared with the MMSE-based Method:} Despite utilizing a significantly longer pilot length and unconstrained feedback, the centralized MMSE-based method performs notably worse than the proposed PCEnet. For instance, at a test SNR of 10 dB, the side channel NMSE of the MMSE-based method is $-$5.07 dB with a pilot length of 16, whereas the proposed PCEnet, even with an imperfect main channel (${\bf h}_{\rm m}$) achieves $-$11.12 dB  using only a pilot length of 4. }
    \item {\bf Notable Improvement in Side Channel Acquisition compared with E2E-AI4CSI:} For example, at 6 dB SNR, the proposed method achieves an NMSE of $-9.94$ dB with a pilot length of 4, compared to $-9.59$ dB for the E2E-AI4CSI NN with a pilot length of 8. This means the proposed method reduces the channel acquisition overhead (including pilot and feedback) by half without sacrificing performance. In contrast, the ``Direct Mapping'' method \cite{9048929,10423003} that infers the side channel directly from the main channel performs poorly, achieving an NMSE of only $-0.14$ dB, indicating its ineffectiveness.
    { The poor performance of direct mapping arises from the BS distance of approximately 160 meters, making it impractical to extrapolate one channel from another under these conditions.}

    \item {\bf Pronounced Performance Improvement at Lower SNRs:}  At $-4$ dB SNR, the proposed method achieves $-4.18$ dB NMSE, outperforming the E2E-AI4CSI's $-0.40$ dB. Similarly, at 10 dB SNR, the proposed method achieves $-11.13$ dB compared to the E2E-AI4CSI's $-10.27$ dB. In scenarios with high channel estimation noise (e.g., low SNRs), the pilot signal carries limited information, so position information becomes more valuable.

    \item {\bf Gap between Perfect and Imperfect Main Channel Localization:} There is a 2 dB accuracy gap between methods utilizing perfect and imperfect main channel estimates, highlighting the potential of PCEnet when more advanced localization methods are employed. {Notably, recent localization studies \cite{9797028,10522995} have achieved a mean localization error of 2.84 meters, particularly with the incorporation of multi-modal information. This underscores the practical applicability of the proposed PCEnet framework in real-world deployments.}
    
\item {{\bf Robustness  to Different SNRs:} Compared to AI-based methods, the MMSE-based method exhibits a smaller performance degradation when the test SNR decreases. For example, when the test SNR drops from 10 dB to $-$4 dB, the NMSE increases by 6.96 dB for PCEnet (with an imperfect main channel ${\bf h}_{\rm m}$, using a pilot length of 4), and by 11.27 dB for E2E-AI4CSI (with a pilot length of 8). In contrast, the MMSE-based method experiences a performance drop of only 3.58 dB.
This difference arises because the AI models are trained at a fixed SNR of 10 dB, leading to a mismatch between training and test SNRs, whereas the MMSE-based method remains unaffected by such discrepancies. A potential solution to mitigate this issue could be training with a mixture of SNRs to enhance generalization. }
\end{itemize}

\subsection{Performance of the One-Sided Real-Time PCEnet Framework}
Figure \ref{oneSided} illustrates the NMSE performance of the proposed one-sided real-time PCEnet for the side channel. In this approach, the estimated position information is solely used in the side channel reconstruction module without introducing any modifications to the pilot design or pilot signal compression. As a result, the full potential of the primary PCEnet is not realized in this scenario.

As expected, the one-sided method does not perform as well as the primary PCEnet method. However, it still significantly outperforms the E2E-AI4CSI framework, which does not incorporate position information. For instance, at an SNR of 6 dB, the NMSE achieved by the one-sided method, with a pilot length of 4, is $-7.79$ dB, compared to $-7.93$ dB for the E2E-AI4CSI framework with a pilot length of 6. This indicates that the one-sided method has the potential to reduce the pilot and feedback overhead for side channel acquisition by approximately one-third. Furthermore, similar to the results from the previous section, feedback accuracy can be further improved if the position is estimated based on the perfect main channel.

\subsection{Performance of the Position Label-Free PCEnet Framework}

In this section, we first evaluate the performance of relative position generation using the main-channel-to-side-channel characteristics mapping. Following that, we present and analyze the overall accuracy of the position label-free PCEnet framework.

To illustrate the results more clearly, we randomly select three groups of UEs, as shown in Figure \ref{OriginalLOC}. In each group, the UEs are positioned close to each other in a column, and each UE is represented as a point on the plot. UEs in the same group are assigned the same color to distinguish between the groups visually. 
Figures \ref{CCae0} and \ref{CCae} show the relative positions of the UEs generated by the vanilla autoencoder and the proposed main-channel-to-side-channel characteristics mapping, respectively. Since the focus is on relative positioning, Figure \ref{CCresult} omits axis labels for simplicity. In Figure \ref{CCae0}, UEs within the same group, such as the red points, are spread across a wide area, indicating that the vanilla autoencoder does not accurately capture the relative position relationships. In contrast, Figure \ref{CCae} demonstrates that the UEs within each group are positioned closely together when using the main-channel-to-side-channel characteristics mapping. This results in a more compact and accurate representation of their relative positions, effectively demonstrating the improvement in capturing position relationships.

\begin{figure}[t]
    \centering
    \includegraphics[width=0.98\linewidth]{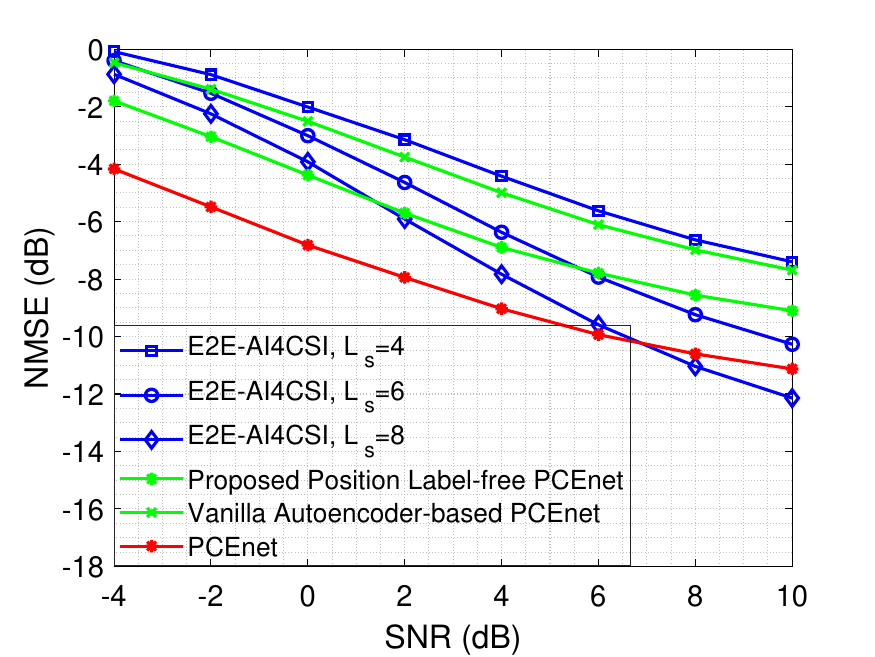}
    \caption{NMSE (dB) performance of the proposed position label-free channel extrapolation for the side channel. The pilot length, $L_{\rm s}$, for the primary PCEnet and its two variants are all set to 4.}
    \label{positionFree}
\end{figure}

Next, Figure \ref{positionFree} presents the final accuracy of the position label-free PCEnet framework. In this figure, the method labeled as ``Vanilla Autoencoder-based PCEnet'' uses the latent representation generated by a vanilla autoencoder as relative position information to assist in side channel acquisition. Despite not utilizing the UEs' actual positions, this method achieves performance comparable to the E2E-AI4CSI framework. This is consistent with the findings from Figure \ref{CCresult}, which highlight the limitations of the vanilla autoencoder in efficiently capturing position information.

By incorporating the main-channel-to-side-channel characteristics mapping, we observe a significant improvement in side channel acquisition performance. For example, at low SNRs, such as 0 dB, the proposed position label-free method with a pilot length of 4 outperforms the E2E-AI4CSI method with a pilot length of 8, effectively reducing channel acquisition overhead by half. Even at higher SNRs, the proposed method achieves performance comparable to the E2E-AI4CSI method with a pilot length of 6, further reducing the acquisition overhead by one-third. These results demonstrate the effectiveness of the proposed position label-free approach, which not only improves performance but also reduces system overhead.

\section{Conclusion {and Future Directions}}
\label{s5}

In this study, we introduced a novel deep learning-based position-domain channel extrapolation framework for cell-free massive MIMO, designed to reduce channel acquisition overhead. Our approach distinguishes itself by leveraging the UE's position as a bridge between the main and side channels. Specifically, it utilizes the acquired main channel information to aid in the estimation and feedback of the side channels. Unlike current learning-based end-to-end channel acquisition methods, our approach enables the BS to infer the UE's position after acquiring the main channel and use this estimated position to guide pilot design and channel reconstruction for the side channel. This effectively reduces the pilot and feedback overhead for the side channels by 50\%.

Additionally, we addressed two practical challenges in implementing the proposed method. First, we retained the existing pilot design and signal compression procedures while incorporating only the estimated position from the main channel into the side channel reconstruction module. This approach improves acquisition accuracy while maintaining low latency. Second, recognizing the challenge of collecting position labels, we proposed a position label-free channel extrapolation strategy, which infers the UE's relative position using a main-channel-to-side-channel characteristics mapping. Simulation results demonstrate that even under highly constrained scenarios, these two strategies can still significantly reduce the pilot and feedback overhead by approximately one-third.

While the proposed learning-based position-domain channel extrapolation shows promising results for cell-free massive MIMO systems, several challenges merit further in-depth investigation in future research:
\begin{itemize}
    \item {\bf Extension to Diverse MIMO Scenarios:} Future research should explore additional promising MIMO scenarios, such as RIS-assisted communications, beyond cell-free massive MIMO, to address specific challenges and capitalize on their distinctive structural benefits.
    \item {\bf Optimization for Specific Working Modes:} The proposed method, compatible with both FDD and TDD systems, should be further optimized for their distinct characteristics (e.g., TDD's channel reciprocity) to reduce CSI acquisition overhead.
    \item {\bf Interoperable Training between BS and UE:} NNs are deployed at both the BS and UE, yet sharing NN information between the BS and UE is often restricted due to intellectual property concerns. A practical training protocol should be developed for the proposed PCEnet, enabling collaboration without requiring NN exchange between the BS and UE.
\end{itemize}

\bibliographystyle{IEEEtran}
\bibliography{reference}

\end{document}